\documentstyle[preprint,aps]{revtex}
\begin{document}
\draft
\preprint{SOGANG-HEP 207/96}

\title{Topological Massive Gauge Theories in Three Dimensions Based on 
the Faddeev-Jackiw Formalism}
\author{Honglo Lee, Yong-Wan Kim, and Young-Jai Park}
\address{Department of Physics, \\
 Sogang University, C.P.O. Box 1142, Seoul 100-611}
\date{Received 29 August 1996}
\maketitle

\begin{abstract}
We quantize the (2+1)-dimensional self-dual and Maxwell-Chern-Simons theories by
using the Faddeev-Jackiw formulation and 
compare the results with those of the Dirac formalism. 
\end{abstract}

\newpage
\section{Introduction} 
The basic ideas of quantization of a constrained system were first 
presented by Dirac \cite{dirac}. By using his method, one can obtain the 
Dirac brackets, which are the bridges to the commutators in quantum theory. 
Several years ago, Faddeev and Jackiw (FJ) proposed a method of symplectic 
quantization of constrained systems for a first-order Lagrangian 
\cite{faddeev},which was different from the Dirac procedure. In the FJ 
method, the 
classification of constraints as first or second class, primary or 
secondary, is not necessary. All constraints are held to the same 
standard. Since their work, their quantization method has attracted much 
attention
because it seems to be algebraically much simpler than the
Dirac method.

In addition, the study of gauge theories in three-dimensional 
space-time is very attractive. In odd-dimensional space-time, 
the topologically non-trivial, gauge-invariant 
Chern-Simons term gives rise to masses for the gauge fields \cite{siegel}. 
It is 
known that the spin-one theory in 2+1 dimensions may be described by two 
covariant actions; one is the Maxwell-Chern-Simons (MCS) action, which is 
constructed with a Maxwell term and a Chern-Simons term \cite{deser}, while 
the other is the self-dual (SD) 
action generating the square root of the Proca equation for a massive 
vector field \cite{townsend}.

In this paper, we will use both the Dirac and the FJ methods in order to 
quantize the (2+1)-dimensional gauge theories. We will derive the Dirac 
brackets and the equivalent equations of motion for the 
SD model \cite{townsend,ywkim} in Section II, and for the MCS theory 
\cite{deser,jackiw} in Section III. Section IV presents the conclusion.

\section{SD model}

\bf 1. Dirac Quantization of the SD Model
\rm

In this subsection, we first briefly recapitulate the Dirac method with 
the SD Lagrangian, 
which is constructed with both the ordinary and the topological mass terms:
\begin{equation}
 {\cal L} = \frac{1}{2} m^2 B_\mu B^\mu - \frac{1}{2} m\epsilon_{\mu\nu\rho} 
B^\mu \partial^{\nu}B^\rho 
\end{equation}
where $g_{\mu\nu} = diag(1,-1,-1)$ and $\epsilon_{012} = 1$.  Denoting the canonical momenta of 
the vector field as $\Pi_\mu$, we obtain three primary constraints and the canonical Hamiltonian as follows
\begin{eqnarray}
 \omega_0 & \equiv & \Pi_0 \approx 0\nonumber ,\\
 \omega_i & \equiv & \Pi_i + \frac{1}{2}m\epsilon_{ij}B^j \approx 0\nonumber ;~~(i=1,2), 
\end{eqnarray}
\begin{equation}
 H_c = \int d^2x~ [-\frac{1}{2}m^2 B^\mu B_\mu + m\epsilon^{ij} B_0 
\partial_i B_j]. 
\end{equation}
With these primary constraints and the corresponding Lagrange 
multipliers $\lambda^\mu$, we write the primary Hamiltonian as
\begin{equation}
 H_p = H_c + \int d^2x~ \lambda^\mu \omega_\mu ;~~(\mu=0,1,2) . 
\end{equation}
Then, we obtain one more constraint by requiring the time stability of 
$\omega_0$: 
\begin{eqnarray}
 \omega_3 \equiv \dot\omega_0 &=& \{\omega_0, H_p\}\nonumber\\
                              &=& m^2 B^0 - m\epsilon^{ij} \partial_i 
B_j~  \approx~ 0~. 
\end{eqnarray}
The time stabilities of $~\omega_i$ and $\omega_3 ~$ give no 
additional constraints and only play the role of fixing the values of 
Lagrange multipliers. All four constraints are completely second-class constraints.

According to the Dirac formalism, we can find the $C_{\mu\nu}$-matrix 
from the Poisson bracket of the constraints \begin{equation}
 C_{\mu\nu} = \{\omega_\mu, \omega_\nu \} = m \left ( \begin{array}{cccc}
                                               0  &  0  &  0  &  -m \\
                                               0  &  0  & -1  &  
\partial^x_2 \\
                                               0  &  1  &  0  &  
-\partial^x_1 \\
                                               m  &  \partial^x_2 & 
-\partial^x_1 & 0 \\
                                           \end{array}
                                  \right ) \delta^2 (x-y)
\end{equation}
with the inverse matrix
\begin{equation}
 C^{-1}_{\mu\nu} = -\frac{1}{m^2}
           \left (\begin{array}{cccc}
                   0 & \partial^x_1 & \partial^x_2 & -1 \\
                   \partial^x_1 & 0 & -m & 0 \\
                   \partial^x_2 & m & 0  & 0 \\   
                   1 & 0 & 0 & 0 \\
                  \end{array}
           \right ) \delta^2 (x-y) .
\end{equation}
Imposing all the constraints, the reduced Hamiltonian is found to the
\begin{equation}
 H_r = \int d^2x~ [\frac{1}{2}(\epsilon^{ij}\partial_i B_j)^2 - \frac{1}{2} 
m^2 B_i B^i] 
\end{equation}
in which the only physical variables are the $B_i$. On the other hand, 
since the Dirac bracket of two variables is defined as 
\begin{equation}
 \{A,B\}_D =  \{A,B\} - \{A,\omega_\mu\} C^{-1}_{\mu\nu}\{\omega_\nu,B\} ,
\end{equation}
the non-trivial Dirac brackets of the variables in this model are 
\begin{equation}
 \{B_i(x), B_j(y)\}_D = -\frac{1}{m} \epsilon_{ij}\delta^2 (x-y).
\end{equation}

\bf 2. FJ Quantization of the SD Model
\rm

Now, we quantize the SD model following the FJ method 
\cite{faddeev,neto,yjpark}.
The first-order Lagrangian equivalent to the Lagrangian in Eq.(1) is
\begin{equation}
 {\cal L}_{SD} = \frac{m}{2} \epsilon_{ij} B^i \dot B^j + {\cal H}^{(0)}(\xi) 
\end{equation} 
where the zeroth-iterated symplectic potential is
\begin{eqnarray*}
  \hspace{0.25in} {\cal H}^{(0)}(\xi) \equiv m\epsilon_{ij}B^0\partial^i B^j - 
 \frac{1}{2} m^2 B_\mu B^\mu .
\end{eqnarray*}
With the initial set of symplectic variables, $\xi^{(0)i}=(B^0,B^1,B^2)$, 
we have, according to the FJ method, the canonical 
one-form $a^{(0)}_i = (0,-\frac{m}{2} B^2 ,\frac{m}{2} B^1)$. 
These result in the following singular symplectic two-form matrix:
\begin{equation}
 f^{(0)}_{ij}(x,y) = m\left ( \begin{array}{ccc}
                       0 & 0 & 0 \\
                       0 & 0 & 1 \\
                       0 & -1 & 0 \\
                      \end{array}
              \right ) \delta^2(x-y) .
\end{equation}
Note that this matrix has a zero mode, $\tilde v^{(0)}_k(x) = (v_1(x),0,0)$, where 
$v_1(x)$ is an arbitrary function. From this zero mode, we get the following 
constraint $\Omega^{(0)}$: 
\begin{eqnarray}
 0 &=& \int d^2x~ \tilde v_k^{(0)} \frac{\delta}{\delta\xi^{(0)k}(x)} 
\int d^2y~ {\cal H}^{(0)}(\xi) \nonumber\\[.15in]
   &=& \int d^2x~ v_1(x) ~[m\epsilon_{ij}\partial^i B^j - m^2 B^0] 
\nonumber\\[.15in]
   & \equiv & \int d^2x~ v_1(x) \Omega^{(0)}~ .
\end{eqnarray}
In order to provide a consistent description of the system for this constraint, the constrained manifold must be stable under time 
evolution. In fact, this constraint is stable under time evolution.

According to the FJ method, we can write the first-iterated Lagrangian 
with a new Lagrange-multiplier as follows:
\begin{equation}
 {\cal L}^{(1)} = \frac{m}{2} \epsilon_{ij} B^i \dot B^j - \frac{m}{2} B^2 
\dot B^1 + 
 \Omega^{(0)}\dot\alpha - {\cal H}^{(1)}(\xi)
\end{equation}
where the first-iterated symplectic potential is
\begin{eqnarray}
{\cal H}^{(1)}(\xi) = \frac{1}{2} (mB^0)^2 - \frac{1}{2} m^2 B_i B^i. 
\end{eqnarray}
Then, the first-iterated set of symplectic variables becomes 
$\xi^{(1)i}=(B^0,B^1,B^2,\alpha)$, and 
the canonical one-form becomes $a^{(1)}_i = (0, -\frac{m}{2} B^2, \frac{m}{2} B^1, 
m\epsilon_{ij} \partial^i B^j - m^2 B^0 )$. We get the following 
first-iterated symplectic matrix from the above variables:
 \begin{equation}
 f^{(1)}_{ij}(x,y) = m\left(\begin{array}{cccc}
                       0 & 0 & 0 & -m \\
                       0 & 0 & 1 & -\partial^2_x \\
                       0 & -1 & 0 & \partial^1_x \\              
                       m & -\partial^2_x & \partial^1_x & 0 \\
              \end{array}
                \right) \delta^2(x-y) .
\end{equation}
Since this is a non-singular matrix, we finally obtain the desired inverse matrix of 
the above matrix as
\begin{equation}
 [f^{(1)}_{ij}]^{-1}(x,y) = \frac{1}{m^2}
                            \left( \begin{array}{cccc}
            0 & -\partial^1_x &  -\partial^2_x & 1 \\
            -\partial^1_x & 0 & -m & 0 \\
            -\partial^2_x & m & 0 & 0 \\
            -1 & 0 & 0 & 0 \\ 
                           \end{array}
                        \right) \delta^2(x-y) .
\end{equation}
Finally, according to the FJ method, the Dirac brackets are acquired 
directly from the elements of the inverse of the symplectic matrix because
\begin{equation}
 \{\xi^{(1)}_i (x) , \xi^{(1)}_j (y)\} = [f^{(1)}]^{-1}_{ij} (x,y).
\end{equation}
Reading the Dirac brackets from above matrix, we find that 
\begin{equation}
 \{B_i(x),B_j(y)\}_D = -\frac{1}{m} \epsilon_{ij}\delta^2(x-y),
\end{equation}
which is the same as the equation for the Dirac brackets in Eq.(9). In 
addition, using the constraints and the Dirac brackets, we 
can easily obtain the self-dual equation of motion for $B_1$, which 
has only one dynamical degree of freedom, 
\begin{equation}
 (\Box + m^2) B_1 = 0 , 
\end{equation}
because $\Pi_1$ is propotional to $B_2$.

It seems appropriate to comment on the Dirac and the FJ formalisms.
Firstly, through the quantization of the SD model, we have shown that the 
number of constraints is fewer and 
the structure of these constraints is very simple because we do not need 
to distinguish between first- or 
second-class constraints, primary or secondary constraints, etc..  Secondly, we have 
easily obtained the Dirac brackets by reading them directly from the inverse 
matrix 
$f^{ij}(x,y)$ of the symplectic two-form matrix. Thirdly, we have shown that 
the symplectic Hamiltonian at the final stage of iterations exactly gives 
the reduced physical Hamiltonian, which may be obtained through several 
steps with the three definitions of the canonical, the total, and the 
reduced 
Hamiltonians in the usual Dirac formulation for constrained systems. 

The above three merits have been recently analyzed in several papers, 
on the subjects of the nonrelativistic point particle, 
three-dimensional topologically massive electrodynamics, the nonlinear 
sigma model, two-dimensional induced gravity\cite{neto}, constrained 
systems\cite{yjpark}, etc. These works show how efficient the 
symplectic formalism is, and confirm that the symplectic 
quantization method is a simpler alternative to 
the Dirac's formalism in the sense that the brackets are obtained more 
easily and are exactly same as the Dirac brackets.
As a result, we can replace the obtained brackets with the quantum 
commutators as $\{~ ,~ \}_D \rightarrow i[~ , ~]$.

\section{MCS theory}
\bf 1. Dirac Quantization of the MCS Theory
\rm

In this subsection, in order to compare it with the FJ formalism, we 
sketch the Dirac quantization procedure with the MCS theory, which is 
constructed with the Maxwell and the topological mass terms: 
\begin{equation}
 {\cal L}_{MCS} = -\frac{1}{2} F^\mu F_\mu + \frac{1}{2} m F^\mu A_\mu
\end{equation}
where $F_\mu \equiv \frac{1}{2}\epsilon_{\mu\nu\rho}F^{\nu\rho} = 
\epsilon_{\mu\nu\rho}\partial^\nu A^\rho$. 
Denoting the canonical momenta of the vector field as $\Pi_\alpha$, we obtain one 
primary constraint and the canonical Hamiltonian as follows:
\begin{equation}
 \omega_0 = \Pi_0 \equiv 0 ,
\end{equation}
\begin{eqnarray}
 H_c = \int d^2x&[& -\frac{m}{2} \epsilon_{ij} \Pi_i A^j + \frac{1}{2} 
(\Pi_i)^2 + \frac{m^2}{8} (A^i)^2 + \frac{1}{2} (\epsilon_{ij} \partial^i A^j)^2 \nonumber\\
     && + ~\partial_i \partial^i A^0 - \frac{m}{2} \epsilon_{ij} 
(\partial^i A^j) A_0]~. 
\end{eqnarray}   
With the primary constraint and the corresponding Lagrange multiplier 
$u$, we write the primary Hamiltonian as
\begin{equation}
  H_p =  H_c +  \int d^2x~ u\omega_0.
\end{equation}
Requiring time stability of the  primary constraint, we get 
one more constraint:
\begin{equation}
 \omega_1 \equiv \dot\omega_0 = \partial^i\Pi_i + \frac{m}{2} 
\epsilon_{ij} \partial^i A_j.
\end{equation}
Note that the time stability of $\omega_1$ gives no additional 
constraints and only plays the role of fixing the value of the Lagrange 
multiplier. These two constraints are first class, which gives rise to 
gauge invariance. Therefore, we should introduce a gauge-fixing 
function to find the true physical variables correctly. 
Choosing the Coulomb gauge condition $\omega_2 = \partial_i A^i$, we 
obtain one more constraint: 
\begin{equation}
 \omega_3  \equiv  \dot \omega_2 = m\epsilon_{ij} \partial^i A^j + \partial_i\partial^i A^0 .
\end{equation}
Now, all four constraints are second-class.

We find the $C_{\mu\nu}$-matrix from the Poisson bracket of the 
constraints:
\begin{equation}
 C_{\mu\nu} = \nabla^2
      \left ( \begin{array}{cccc}
      0 & 0 & 0 & -1 \\
      0 & 0 & 1 & 0  \\
      0 & -1 & 0 & 0 \\
      1 & 0 & 0 & 0  \\
      \end{array}
     \right) \delta^2(x-y)
\end{equation}
with its inverse
\begin{equation}
 C^{-1}_{\mu\nu} = \frac{1}{\nabla^2}
          \left ( \begin{array}{cccc}
           0 & 0 & 0 & 1 \\
           0 & 0 & -1 & 0  \\
           0 & 1 & 0 & 0 \\
          -1 & 0 & 0 & 0  \\
      \end{array}
     \right) \delta^2(x-y).
\end{equation}
Imposing all the constraints on Eq. (23), the reduced Hamiltonian is 
found to be
\begin{equation}
 H_r = \int d^2x~ [ -\frac{m}{2} \epsilon_{ij} \Pi_i A^j + \frac{1}{2} 
(\Pi_i)^2 +
\frac{m^2}{8} (A^i)^2 + \frac{1}{2} (\epsilon_{ij} \partial^i A^j)^2 ] .
\end{equation}
Through a similar procedure as in the previous section, we obtain the 
following Dirac brackets: 
\begin{eqnarray}
 \{\Pi_i(x),\Pi_j(y) \}_D &=& -\frac{m}{2}\epsilon_{ij}\delta^2(x-y) 
,\nonumber\\
 \{A^i(x),\Pi_j(y) \}_D &=& \frac{\epsilon^{ik}\epsilon_{jl} \partial^k_x 
\partial^x_l}{\nabla^2} \delta^2(x-y) .
\end{eqnarray}
These Dirac brackets will be compared with the symplectic brackets in the next 
subsection. 

\bf 2. FJ Quantization of the MCS Theory
\rm

Now, we quantize the MCS theory following the FJ method. The first-order 
Lagrangian is \begin{equation}
 {\cal L}_{MCS} = \Pi_i \dot A^i - {\cal H}^{(0)}(\xi)
\end{equation}
where the zeroth-iterated symplectic potential is
\begin{eqnarray}
 {\cal H}^{(0)}(\xi) &=& -\frac{m}{2} \epsilon_{ij} \Pi_i A^j + \frac{1}{2}(\Pi_i)^2 +
       \frac{m^2}{8}(A^i)^2 + \frac{1}{2}(\epsilon_{ij}\partial^i A^j)^2 
\nonumber\\
  && + \Pi_i \partial^i A^0 - \frac{m}{2} \epsilon_{ij} (\partial^i A^j) A_0 .
\end{eqnarray}
With the initial set of symplectic variables, $\xi^{(0)i} =(A^0,A^1,A^2,\Pi_1,\Pi_2)$, we 
have, according to the FJ method, the canonical one-form $a^{(0)}_i = 
(0,\Pi_1,\Pi_2,0,0)$. These result in the following singular symplectic 
two-form matrix: 
\begin{equation}
 f^{(0)}_{ij}(x,y) = \left( \begin{array}{ccccc}
 0 & 0 & 0 & 0 & 0 \\
 0 & 0 & 0 & -1& 0 \\
 0 & 0 & 0 & 0 & -1 \\
 0 & 1 & 0 & 0 & 0 \\
 0 & 0 & 1 & 0 & 0 \\
 \end{array}
                \right) \delta^2(x-y).
\end{equation}
This matrix has a zero mode $\tilde v^{(0)}_k(x) = (v_1(x),0,0,0,0)$, where 
$v_1(x)$ is an arbitrary function. Using this zero mode, we get the 
following constraint: 
\begin{eqnarray}
 0 &=& \int d^2x~ \tilde v^{(0)}_k(x) \frac{\delta}{\delta\xi^{(0)k}(x)} 
\int d^2y ~{\cal H}^{(0)}(\xi) \nonumber\\
   &=& -\int d^2x~ v_1(x) [\partial^i \Pi_i + \frac{m}{2} \epsilon_{ij} 
\partial^i A^j] \nonumber\\
   & \equiv & -\int d^2x~ v_1(x) \Omega^{(0)}.
\end{eqnarray}

We can write the first-iterated Lagrangian with a new 
Lagrange-multiplier as
\begin{equation}
 {\cal L}^{(1)}_{MCS} = \Pi_i \dot A^i + \Omega^{(0)} \dot\alpha - 
{\cal H}^{(1)}(\xi) 
\end{equation}
where the first-iterated symplectic potential is
\begin{eqnarray}
 {\cal H}^{(1)}(\xi) &=& {\cal H}^{(0)}(\xi)\mid_{\Omega^{(0)}=0}\nonumber\\
              &=& -\frac{m}{2} \epsilon_{ij} \Pi_i A^j + 
\frac{1}{2}(\Pi_i)^2 +
                    \frac{m^2}{8}(A^i)^2 + \frac{1}{2}(\epsilon_{ij}
\partial^i A^j)^2 .
\end{eqnarray}
Then, the first-iterated set of symplectic variables becomes $\xi^{(1)i} = 
(A^1,A^2,\Pi_1,\Pi_2 , \alpha)$, and the canonical one-form becomes 
$a^{(1)}_i =
(\Pi_1,\Pi_2,0,0,\partial_i\Pi^i + \frac{m}{2} \epsilon_{ij}\partial^i A^j)$. From 
these variables, we find the following first-iterated symplectic matrix:
\begin{equation}
 f^{(1)}_{ij}(x,y) = \left( \begin{array}{ccccc}
 0 & 0 & -1 & 0 & -\frac{m}{2} \partial^2_x \\
 0 & 0 & 0 & -1& \frac{m}{2} \partial^1_x \\
 1 & 0 & 0 & 0 & \partial^1_x \\
 0 & 1 & 0 & 0 & \partial^2_x \\
 -\frac{m}{2} \partial^2_x & \frac{m}{2} \partial^1_x & \partial^1_x & 
\partial^2_x & 0 \\
 \end{array}
 \right) \delta^2(x-y)
\end{equation}
This matrix is also singular. 

Although we use the zero mode, $\tilde
v^{(1)}_k = (\partial_1 v_5,\partial_2 v_5, \frac{m}{2}
\partial_2 v_5,$ $-\frac{m}{2} \partial_1 v_5, v_5)$, which gives 
non-dynamical relations of the system in the FJ method, we can't obtain 
the constraint any more. Since the Lagrangian one-form is invariant under 
the transformation rule of the symplectic variable, 
$\delta\xi^{(1)i} = \tilde v^{(1)}_k \eta$, we should introduce a 
gauge-fixing function. Using the Coulomb gauge condition $\Omega^{(1)} = 
\partial^i A^i$, we can extend the system as follows 
\begin{equation}
 {\cal L}^{(2)}_{MCS} = \Pi_i \dot A^i + \Omega^{(0)} \dot\alpha +   
\Omega^{(1)} \dot\beta - {\cal H}^{(2)}(\xi) 
\end{equation}
where
\begin{eqnarray*}
 {\cal H}^{(2)}(\xi) = {\cal H}^{(1)}(\xi)\mid_{\Omega^{(1)}=0} .
\end{eqnarray*}
The symplectic variables and the canonical one-form of the second-iterated 
Lagrangian are
 \begin{eqnarray}
 \xi^{(2)i} &=& (A^1,A^2,\Pi_1,\Pi_2 , \alpha, \beta) ,\nonumber\\
 a^{(2)}_i &=& (\Pi_1,\Pi_2,0,0,\partial_i\Pi^i + \frac{m}{2}\epsilon_{ij}\partial^i A^j, 
 \partial_i A^i )~. 
\end{eqnarray}
Then, the symplectic two-form matrix is
\begin{equation}
 f^{(2)}_{ij}(x,y) = \left( \begin{array}{cccccc}
 0 & 0 & -1 & 0 & -\frac{m}{2} \partial^2_x & -\partial^1_x \\
 0 & 0 & 0 & -1& \frac{m}{2} \partial^1_x & -\partial^2_x \\
 0 & 0 & 0 & 0 & \partial^1_x & 0 \\
 1 & 0 & 0 & 0 & \partial^2_x & 0 \\
 -\frac{m}{2} \partial^2_x & \frac{m}{2} \partial^1_x & \partial^1_x & \partial^2 & 0 & 0 \\
 -\partial^1_x & -\partial^2_x & 0 & 0 & 0 & 0 \\
 \end{array} \right) \delta^2(x-y) .
\end{equation}
Since this matrix is non-singular, we finally obtain inverse as
\begin{equation}
 [f^{(2)}_{ij}]^{-1}(x,y) = \frac{1}{\nabla^2} \left( \begin{array}{cccccc}
 0 & 0 & \partial^2_x \partial^x_2 & -\partial^1_x \partial^x_2 & 0 & \partial^1_x \\
 0 & 0 & -\partial^1_x \partial^x_2 & \partial^1_x \partial^x_1 & 0 & \partial^2_x \\
 \partial^2_x \partial^x_2 & \partial^1_x \partial^x_2 & 0 & -\frac{m}{2}&\partial^1_x &
\frac{m}{2} \partial^2_x \\
 \partial^1_x \partial^x_2 & \partial^1_x \partial^x_1 & \frac{m}{2} & 0 & 
-\partial^2_x & -\frac{m}{2} \partial^1_x  \\
 0 & 0 & -\partial^1_x & -\partial^2_x & 0 & 1 \\
 \partial^1_x & \partial^2_x & \frac{m}{2} \partial^2_x & -\frac{m}{2} 
\partial^1_x & -1 & 0 \\
 \end{array}
                \right) \delta^2(x-y) .
\end{equation}
Then, we can directly read the Dirac brackets for the true 
physical fields from  the above matrix, and they are the same as those in 
Eq. (29). In adobtion, we know that the physical degree of freedom in the configuration 
space is only one. The equation of motion of this degree of freedom, which is really the dual 
field $F^0$ contained in the Lagrangian in Eq. (20),  is obtained by using 
the Dirac brackets in Eq. (29) and is found to be 
\begin{equation}
 (\Box + m^2) F^0 = 0 .
\end{equation}
Therefore, the field $F^0$ is effectively equivalent to the field $B_1$, 
with the same mass appearing in both Eqs. (19) and (41).


\section{Conclusion} 
In conclusion we have studied the MCS theory and the SD gauge theory in 
(2+1)-dimensions using the Dirac and the FJ fomulations.
We have found that both the Dirac and the FJ formulations result in 
the same Dirac brackets. Especially, we ascertain that for these cases 
the FJ formulation also is algebraically a much simpler method, which  
gives the
desired Dirac brackets readily without the classification of constraints, than 
that of Dirac's just as several other interesting models \cite{neto,yjpark}.
We have also shown that both the MCS theory and the SD gauge theory have 
only one degree of freedom in the configuration space and have effectively 
the same equations of motion. From this fact, we have found through FJ 
quantization that both 
theories are equivalent to each other at the level of the equation of 
motion. This result 
coincides with that of Faddeev and Jackiw obtained by using the Master 
equation \cite{jackiw}.

\acknowledgments

  The present study was supported by the Basic Science Research Institute
Program, Ministry of Education, Project No. BSRI-96-2414.


\begin{references}
\bibitem{dirac} P. A. M. Dirac, \it Lectures on Quantum Mechanics \rm 
(Belfer Graduate School, Yeshiba University Press, New York, 1964).
\bibitem{faddeev}L. Faddeev and R. Jackiw, Phys. Rev. Lett. {\bf 60}, 
1692  (1988).
\bibitem{siegel} W. Siegel, Nucl. Phys. {\bf B156}, 135 (1979); R. Jackiw 
and S. Templeton, Phys. Rev. {\bf D23}, 2291 (1981).
\bibitem{deser} S. Deser, R. Jackiw, and S. Templeton, Ann. Phys. {\bf
140}, 372 (1982).
\bibitem{townsend} P. K. Townsend, K. Pilch, and P. van  Nienwenhuizen, 
Phys. Lett. {\bf B136}, 38 (1984).
\bibitem{ywkim} Y. W. Kim, Y. J. Park, K. Y. Kim, and Y. Kim, Phys. Rev. 
{\bf D51}, 2943 (1995).
\bibitem{jackiw} S. Deser and R. Jackiw, Phys. Lett. {\bf B139}, 371 (1984).
\bibitem{neto} M. M. Horta Barreira and C. Wotzasek, Phys. Rev. {\bf D45}, 
1410 (1992); J. Barcelos-Neto and C. Wotzasek, Mod. Phys. Lett. {\bf A7}, 1737 (1992); Int. J. Mod. Phys. 
{\bf A7}, 4981 (1992); C. Wotzasek and C. Neves, J. Math. Phys. {\bf 34}, 
1807 (1993); C. Han, Phys. Rev. {\bf D47}, 5521 (1993).
\bibitem{yjpark} D. S. Kulshreshtha and H. J. W. M$\ddot u$ller-Kirsten, 
Phys. 
Rev. {\bf D45}, R393 (1992); N. Banerjee, D. Chattergee, and S. Ghosh, 
Phys. Rev. {\bf D46}, 5590 (1992); Y-W. Kim, Y-J. Park, K. Y. Kim, Y. Kim, 
and 
C-H. Kim, J. Korean Phys. Soc. {\bf 26}, 243 (1993); J. W. Jun and C. Jue, 
Phys. Rev.  {\bf D50}, 2939 (1994); S-J. Yoon, Y-W. Kim, S-K. Kim, Y-J. 
Park, K.Y. Kim, and Y. Kim, J. Korean Phys. Soc. {\bf 27}, 270 (1994); Y-W. 
Kim, Y-J. Park, K.Y. Kim, and Y. Kim, J. Korean Phys. Soc. {\bf 27}, 610 
(1994); Y-W. Kim, Y-J. Park, and Y. Kim, J. Korean Phys. Soc. {\bf 28}, 773 
(1995); E-B. Park, Y-W. Kim, Y-J. Park, and 
Y. Kim, Mod. Phys. Lett. {\bf A10}, 1119 (1995). 
\end{references}
\end{document}